# Clouds effect on the Atmospheric Total Column Carbon Dioxide Retrieval by Space Orbiting Argus 1000 Micro-spectrometer: Introductory Study

**Naif Alsalem[1], Rehan Siddiqui[1], Catherine Tsouvaltsidis[2], and Brendan Quine[1,2]**

[1]Physics and Astronomy Department, York University, 4700 Keele St, Toronto, ON M3J 1P3; (naif11, rehanrul)@yorku.ca.

[2]Earth & Space Science and Engineering Department, York University, 4700 Keele St, Toronto, ON M3J 1P3; (tsouvalc, bquine)@yorku.ca

Corresponding author: Naif Al Salem (naif11@yorku.ca, al-naif1@hotmail.com)

**Abstract**

Carbon Dioxide ($CO_2$) is one of the most important greenhouse gases after water vapor ($H_2O$) which plays significant role in the climate process. Accurate space-based measurement of $CO_2$ is of great significance in inferring the location of $CO_2$ sources and sinks. Uncertainties in greenhouse gases (GHG) retrieval process must be minimized to accurately infer the actual amount of the atmospheric species. Clouds pose a large uncertainty in $CO_2$ space-based retrieval process leading, mostly, to an underestimation in the $CO_2$ absorption amount above the cloud layer provided that photons do not perform multiple paths. In this paper, three different cases of data collected over cloudy and clear skies by Argus 1000 micro-spectrometer were analyzed. Findings show that the $CO_2$ absorption in the absence of clouds is approximately 4.5% higher than when clouds are present.

## 1 Introduction

This introductory analysis aims to explore and analyze the data collected by the Argus 1000 micro-spectrometer onboard the CanX-2 Nanosatellite. Specifically, it focuses on the effect of clouds on the $CO_2$ absorption in the short wavelength infrared (SWIR) range. Clouds are one of the most variable and significant parameters in climate studies as they reflect some fraction of the incident solar radiation and also absorb a significant portion of the long wavelength infrared (LWIR) that is emitted from the Earth's surface and atmosphere.

Clouds cover approximately 65% of the Earth and are the most important regulator of solar radiation. They reflect the incoming solar radiation back to space and therefore cooling the Earth–atmosphere system. The cloud reflectivity (albedo) depends on cloud type, form and solar zenith angle. Clouds also absorb solar radiation in the near infrared region [Fu Q, 2002]. They appear to absorb up to 20% of the solar energy incident on them, with solar heating rates reaching over 2 K h$^{-1}$ near cloud tops [Slingo & Schrecker, 1982]. The previous numbers are not definitive, however, and our understanding of cloud absorption remains limited from both observational and theoretical perspectives [Davies *et al*., 1984]. Cloud presence in the atmospheric boundary layer could lead to an overestimate or underestimate of $CO_2$ absorption at 1580 nm and 1600 nm. David *et al*. [1985] showed the high reflection spectra of solar radiation from clouds in the spectral range from 300 nm to 2500 nm (Figure 5 in their paper). Krijger *et al*. [2005] developed an algorithm that identifies cloud-free SCIAMACHY observations in order to accurately detect greenhouse gases such as $CH_4$ and $CO_2$ as clouds contaminate the total column of such gases and therefore affect the quality of these data products. The authors showed the solar radiation behavior over cloudy and cloud-free (over snow covered surfaces) conditions. It was clear that less radiation was monitored at 1.6µm band over snow covered surfaces (cloud-free) than when cloudy (Figure 5 in their paper). Normally, the standard $CO_2$ concentration may vary with the surface of reflection or scattering of photons for specified region. Instead, our hypothesis is that the presence of high or low amount of $CO_2$ may affect the finding of cloud scene because of its chemical features [Siddiqui R et al., 2015].

The importance of accurately detecting the greenhouse gases carbon dioxide ($CO_2$) comes from the fact that it is one of the primary atmospheric gases contributing to anthropogenic climate change. Its concentration in the atmosphere has increased by approximately one third since the preindustrial times 1750 [IPCC, 2001] and is now approaching approximately 400 (399.65) parts per million (ppm) [Jones, 2013]. There are, however, many parameters that have a major effect on the $CO_2$ concentration in the atmosphere such as albedo, cloud type, solar zenith angle, elevation, aerosols and water vapor.

O'Brien and Rayner [2002] showed that $CO_2$ absorption in the presence of clouds would lead to an underestimation in $CO_2$ because clouds prevent solar radiation from getting absorbed by $CO_2$ in the underlying atmosphere. However, Mao and Kawa [2004] showed that the aerosols and cirrus clouds would overestimate the column $CO_2$ amount. The reason they stated for enhancing $CO_2$ in the presence of cirrus clouds and aerosols is that the forward scattering is larger than the backscattering when the surface albedo is 0.3 leading to a longer path of the sunlight.

## 2 Materials and Methods

### 2.1 Argus 1000 on CanX-2

The Canadian Advanced Nanospace eXperiment (CanX) is a series of satellites launched by the University of Toronto Institute for Aerospace Studies' Space Flight Laboratory (UTIAS/SFL) since September, 2001 [Rankin, 2005]. The CanX program was developed to provide Canada with a continuous supply of highly skilled and experienced space system and microsatellite engineers while at the same time providing a low-cost, quick-to-launch satellite platform upon which to execute scientific and technology demonstration missions [Sarda *et al.*, 2006]. Argus 1000 micro-spectrometer (Figure 1) is a new generation of miniature remote sensing instruments primarily used to monitor greenhouse gas emissions from space. Launched on the CanX-2 micro-satellite April 28, 2008 [Jagpal *et al.*, 2010], Argus 1000 spectrometer is capable of monitoring ground-based sources and sinks of anthropogenic pollution. The instrument was designed to take nadir observation of the reflected sunlight from Earth's surface and the atmosphere. The nadir viewing geometry mode of Argus is of particular significance as this observation mode provides the highest spatial resolution on the bright land surfaces and is expected to return more useable soundings in regions that are partially cloudy or have significant surface topography. The counts of 138 near infrared channels, corresponding to the spectral range of 900−1700 nm, are logged. From low Earth orbit, an instantaneous spatial resolution of 1.5 km is achieved [Jagpal, 2011].

Traditionally, the Argus 1000 Spectrometer has been used as an atmospheric monitoring spectrometer, collecting data regarding the vertical columns of important atmospheric trace gases such as $CO_2$ and $H_2O$. Other greenhouse gas species such as nitrous oxide ($N_2O$), hydrogen fluoride (HF), methane ($CH_4$) and carbon monoxide (CO) can also be monitored. The instrument can also be used for biological purposes as organic molecules containing aliphatic O-H, C-H, and C-O bonding exhibit absorption bands in the 1700 nm region [Thoth Technology, 2010].

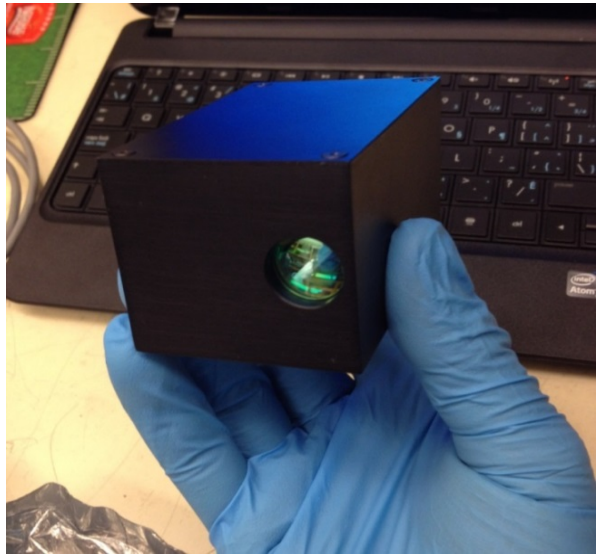

**Figure 1.** Argus 1000 spectrometer commercial unit shown at the Space Engineering Laboratory, York University.

Spectra of reflected radiation from the Earth's surface provide some important absorption features that are associated with the absorption of solar radiation by gases in the atmosphere. Measurements of the absorption of reflected sunlight by $CO_2$ at NIR wavelengths were extremely sensitive to the $CO_2$ concentration change near the surface, where its sources and sinks are located [Boland *et al*, 2009]. The NIR nadir spectra measured by the Argus 1000 spectrometer contain information on the vertical columns of important atmospheric trace gases such as carbon dioxide ($CO_2$) and water vapour ($H_2O$). Other greenhouse gas species such as nitrous oxide ($N_2O$), hydrogen fluoride (HF), methane ($CH_4$) and carbon monoxide (CO) have NIR absorption features within the Argus spectral range; however, they are relatively weak [Jagpal, 2011].

The capacity of infrared spectroscopy aids in predicting soil properties. These properties are important in assessing soil health and fertility and have direct effects on potential agricultural practices and land degradation. Tsouvaltsidis *et al* [2015] built a low cost, commercial off-the-shelf (COTS) Unmanned Arial Vehicle (UAV) system perform remote spectral data collection. They collected spectral data over various soil types at differing moisture contents and over coastlines. This was done in order to test algorithms that determine soil moisture content (SMC) from spectral imagery and other algorithms designed for the geolocation of data from a spaceborne nadir pointing spectrometer by collecting infrared spectral measurements over coastal areas. Their collected data can be utilized as ground truth to verify the effectiveness of the algorithms developed for the spectral instrument. They also designed ArgusE micro-spectrometer in order to ascertain whether it is possible to obtain surface soil moisture content measurements from space using its short-wave infrared detector [Tsouvaltsidis *et al*, 2015].

### 2.2 GENSPECT

GENSPECT is a line-by-line radiative transfer algorithm for absorption, emission, and transmission for a wide range of atmospheric gases. Given information including gas types and amounts, pressure, path length, temperature, and frequency range for an atmosphere, the GENSPECT model computes the spectral characteristics of the gas. GENSPECT employs a new computation algorithm that maintains a specified accuracy for the calculation as a whole by pre-computing where a line function may be interpolated without a reduction in accuracy. The approach employs a binary division of the spectral range, and calculations are performed on a cascaded series of wavelength grids, each with approximately twice the spectral resolution of the previous one. The GENSPECT error tolerances are 0.01%, 0.1%, and 1% which may be selected according to the application [Quine & Drummond, 2002].
GENSPECT has been developed under MATLAB as a toolbox of components. The toolbox includes a library of functions and a library of scripts to illustrate how to carry out example calculations to model optical paths through planetary atmospheres or laboratory instrumentation [Quine & Drummond, 2002]. The GENSPECT algorithm was used in this introductory analysis to model the solar radiance that follows a single path under two different atmospheric conditions (cloudy and cloud-free atmospheres).

### 2.3 Theoretical Effect of Clouds

Two scenarios that the solar radiation might follow are examined in the model. In the first scenario, the solar radiation reaches and reflects off the Earth's surface back to space. In the second scenario, the solar radiation reflects off a cloud layer that is approximately 4 km above the ground. Figure 2 illustrates the two scenarios that the solar radiation could experience in its journey from space through the atmosphere to the Earth's surface and back to space towards the sensor. The solar radiances $I_{dir}$ and $I_{rfc}$ are the direct beam reflected off the ground and the beam reflected off the cloud layer, respectively. Table 1 lists the model parameters used in the solar radiance path scenarios.

In theory, solar beams will have a chance to experience the full atmospheric path when clouds are absent leading to an accurate quantification of greenhouse gases in the Earth's atmosphere. However, they will suffer from short atmospheric path in the presence of clouds leading to an underestimation of greenhouse gas amount in the atmosphere. Multiple reflections below the cloud layer and multi-path above the cloud layer will lead to an overestimation in the greenhouse gas concentrations.

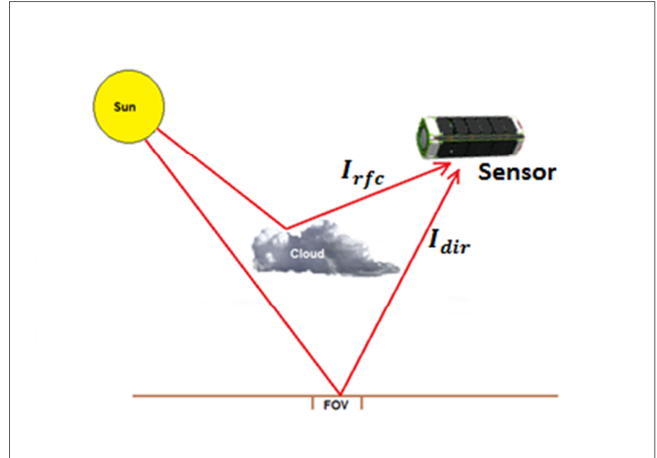

**Figure 2**. Illustration of possible scenarios as solar radiation passes through the atmosphere and back to space to reach the sensor. $I_{dir}$ is the solar radiation beam directly reflected by the Earth's surface toward the sensor field of view (FOV). $I_{crf}$ is the solar beam reflected off clouds.

**Table 1.** Model parameters used in the path scenario calculation.

| Parameter | Value |
| --- | --- |
| 1976 U.S Standard Atmosphere | 0-2 km |
| $CO_2$ Mixing Ratio | 360 ppm |
| Model Resolution | $8\times10^{-3}$ cm$^{-1}$ |
| Altitude | 4 km and Sea level |
| Number of layers | 40 |
| Database Type | Hitran |
| Surface | Lambertian |
| Reflectance (Albedo) | 0.3 (30%) and 0.29 (29%) |
| SZA | 30° |
| Satellite Viewing Angle | Nadir |

### 3 Argus 1000 Data Retrieval

The Argus 1000 spectrometer team at York University is commanding and controlling the instrument alongside CanX-2 satellite operations and the control unit at UTIAS (University of Toronto Institute for Aerospace Studies). Argus collects data over the determined targets for a four-week long period and stops for a two-month break, allowing the other two experiments aboard CanX-2 to function. The Argus team at the Space Engineering Laboratory at York University prepares the observation tables for the desired targets around the globe using the Systems Tool Kit (STK) software. The Argus-1000 target list contains 35 sites around the Earth as shown in Figure 3. Figure 4 shows some of Argus targets and typical CanX-2 overpasses. STK is used to simulate the passes of the CanX-2 satellite over the selected targets providing a list of start and stop times and the duration time in seconds for each pass every week during the observation campaign. Table 2 shows a sample of the typical observation table generated by

STK. The highlighted pass in Table 2 is an indication for the operations team at UTIAS to prioritize it in this campaign. The generated lists of Argus targets are sent to the operations unit at UTIAS to start the observation week. The operations team at UTIAS will then send the collected data to the Argus team at York University for processing and analysis.

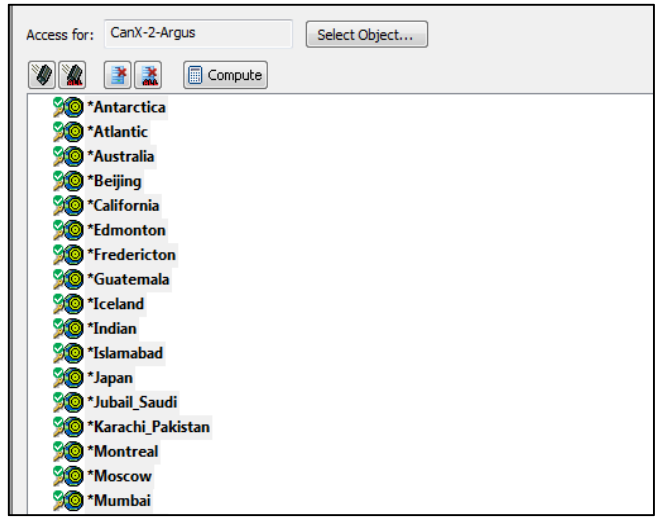

**Figure 3.** Sample of Argus targets in STK.

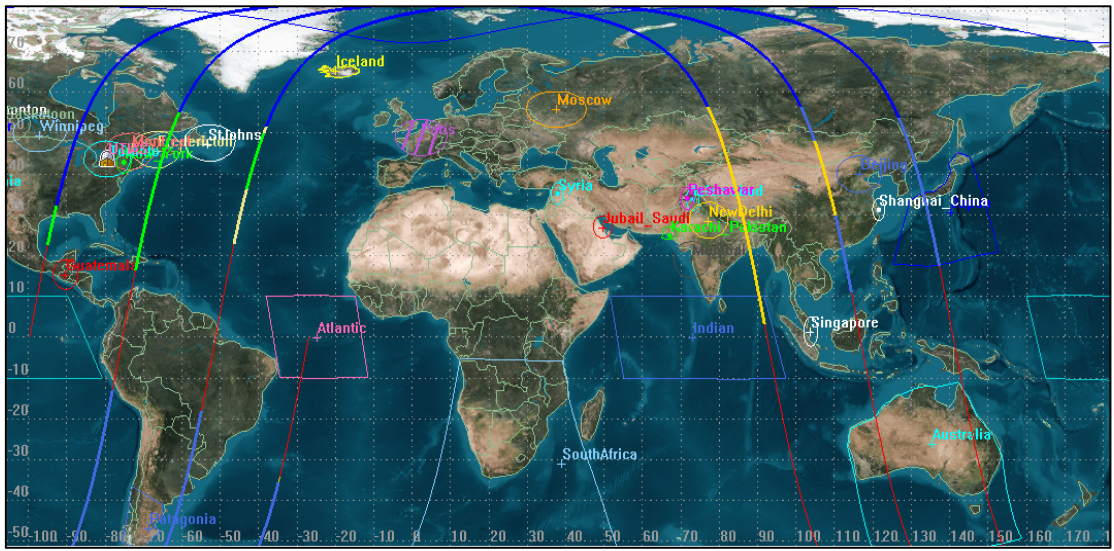

**Figure 4.** Some of Argus targets around the globe and CanX-2 passes over them.

Table 2. List of the selected targets for week 100 of January 19th, 2015.

| Week | Start Time (UTCG) | Stop Time (UTCG) | |
|---|---|---|---|
| 100 | 2015/Jan/19 04:00:00.00 | 2015/Jan/23 04:00:00.00 | |
| Pass | **Start Time (UTCG)** | **Stop Time (UTCG)** | **Duration (sec)** |
| 5 | 2015/Jan/19 10:41:35.35 | 2015/Jan/19 10:44:56.56 | 200.537 |
| 5 | 2015/Jan/19 10:48:06.06 | 2015/Jan/19 10:48:16.16 | 9.36 |
| 5 | 2015/Jan/19 10:54:13.13 | 2015/Jan/19 11:01:32.32 | 438.79 |
| 6 | 2015/Jan/19 12:16:41.41 | 2015/Jan/19 12:20:50.50 | 248.448 |
| 6 | 2015/Jan/19 12:22:54.54 | 2015/Jan/19 12:26:34.34 | 220.573 |
| 6 | 2015/Jan/19 12:30:27.27 | 2015/Jan/19 12:38:29.29 | 482.564 |
| 7 | 2015/Jan/19 13:54:06.06 | 2015/Jan/19 13:56:57.57 | 171.334 |
| 7 | 2015/Jan/19 13:58:37.37 | 2015/Jan/19 14:03:38.38 | 300.831 |
| <span style="color:red">8</span> | <span style="color:red">2015/Jan/19 15:34:30.30</span> | <span style="color:red">2015/Jan/19 15:40:33.33</span> | <span style="color:red">362.207</span> |
| 8 | 2015/Jan/19 15:30:56.56 | 2015/Jan/19 15:32:32.32 | 96.237 |
| 14 | 2015/Jan/20 01:10:14.14 | 2015/Jan/20 01:18:56.56 | 521.904 |
| 14 | 2015/Jan/20 01:28:25.25 | 2015/Jan/20 01:29:42.42 | 76.648 |
| 14 | 2015/Jan/20 01:10:14.14 | 2015/Jan/20 01:18:56.56 | 521.904 |
| 16 | 2015/Jan/20 03:57:12.12 | 2015/Jan/20 03:58:50.50 | 98.466 |
| 16 | 2015/Jan/20 03:57:12.12 | 2015/Jan/20 03:58:50.50 | 98.466 |
| 17 | 2015/Jan/20 05:34:53.53 | 2015/Jan/20 05:36:16.16 | 82.202 |
| 17 | 2015/Jan/20 05:34:53.53 | 2015/Jan/20 05:36:16.16 | 82.202 |
| 17 | 2015/Jan/20 05:34:53.53 | 2015/Jan/20 05:36:16.16 | 82.202 |
| 29 | 2015/Jan/21 00:55:11.11 | 2015/Jan/21 00:56:29.29 | 77.975 |
| 29 | 2015/Jan/21 01:41:33.33 | 2015/Jan/21 01:44:46.46 | 192.859 |
| 31 | 2015/Jan/21 04:13:07.07 | 2015/Jan/21 04:13:28.28 | 20.729 |
| 36 | 2015/Jan/21 12:35:59.59 | 2015/Jan/21 12:39:08.08 | 189.539 |
| 36 | 2015/Jan/21 12:59:14.14 | 2015/Jan/21 13:07:11.11 | 476.287 |
| 52 | 2015/Jan/22 14:03:59.59 | 2015/Jan/22 14:05:40.40 | 100.308 |
| 69 | 2015/Jan/23 17:30:10.10 | 2015/Jan/23 17:32:26.26 | 136.224 |
| 69 | 2015/Jan/23 18:07:35.35 | 2015/Jan/23 18:15:11.11 | 455.572 |

The operations team at UTIAS provides the Argus team with the data sets weekly during the observation campaign. The data sets consist of three files: a binary file that details Argus settings (exposure time, sensitivity, temperature) that were applied during the observation, the data file that has an extension *.CX2MEM containing the raw data collected over the selected target, and the attitude file that provides the satellite orientation information. Figure 5 illustrates the different types of the data files provided by the operations unit at UTIAS. Argus provides the raw data file, which consists of a series of data packets, to an onboard computer in 532 byte unsigned 8-bit words. Argus data packets are transmitted continuously at a cycle period determined as (256 milliseconds + Integration time) * (Number of Scans + 1).

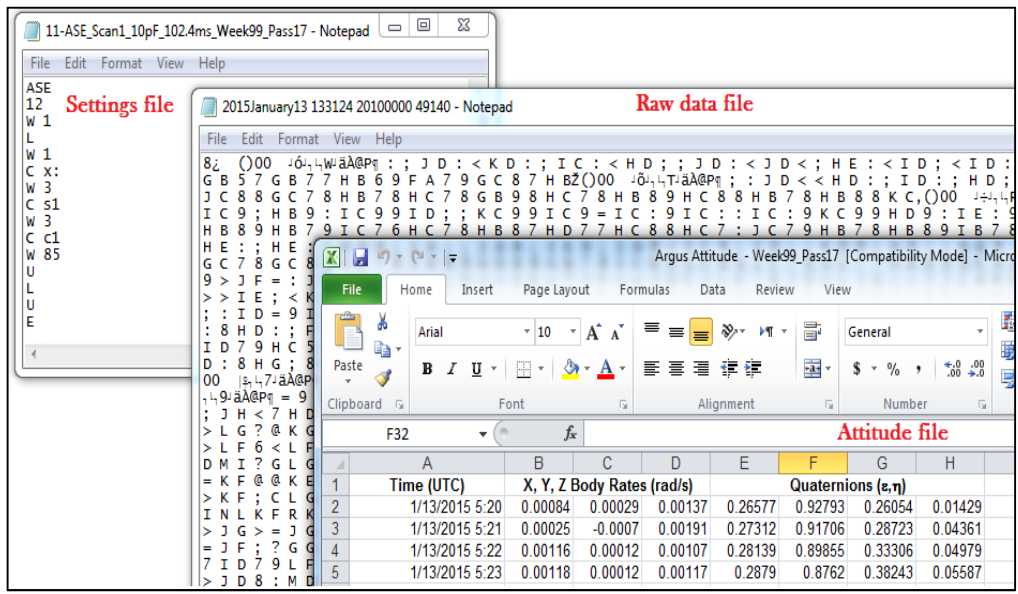

**Figure 5.** Data files provided by UTIAS for each pass in every observational campaign.

Three different cases from three different datasets in which Argus 1000 spectrometer collected cloudy and cloud-free data were analyzed. Cloudy and cloud-free spectra from week 41 pass 34, week 09 pass 36 and week 77 pass 28 were normalized. To investigate the $CO_2$ absorption amount in each case, data were analyzed near the $CO_2$ absorption band 1580 nm by taking the ratio between cloudy and cloud-free spectra. This qualitative introductory analysis will provide and an idea about the $CO_2$ absorption behavior in the absence and presence of clouds. Table 3 lists the week per pass datasets and packet numbers (observation numbers) geolocation parameters

For validation purposes, cloud cover data is retrieved from MODIS (MODerate Resolution Imaging Spectroradiometer) [Savtchenko *et al*, 2003] as well as the AVHRR (Advanced Very High Resolution Radiometer) [SCEC, 2009] to verify whether Argus 1000 spectrometer was collecting data over cloudy scenes.

**Table 3.** Geolocation details for the weeks and passes being analyzed.

| Week/Pass | Packet Number | Date | Coordinates | Status |
|---|---|---|---|---|
| Week 41 pass 34 | 240 | 09/09/2011 | Lat: 31°41'11.40"S  Long: 119°47'59.64"E | Cloud-Free |
| Week 41 pass 34 | 280 | 09/09/2011 | Lat: 34°25'39.00"S  Long: 119° 7'5.16"E | Cloudy |
| Week 09 pass 36 | 26 | 04/11/2009 | Lat: 10°33'34.56"N  Long: 34°34'12.72"W | Cloudy |
| Week 09 pass 36 | 43 | 04/11/2009 | Lat: 8° 4'59.88"N  Long: 35° 4'6.60"W | Cloud-Free |
| Week 77 pass 28 | 66 | 28/08/2013 | Lat: 25°36'22.32"S  Long: 129°56'52.08"E | Cloud-Free |
| Week 77 pass 28 | 70 | 28/08/2013 | Lat: 26°37'45.84"S  Long: 129°42'38.52"E | Cloudy |

## 4 Results and Discussion

### 4.1 Model Effect of Clouds

GENSEPECT was used to create two scenarios that the solar radiation might follow in its journey to the Earth's surface and back to space reaching the sensor. As mentioned in section 2, one of the solar beams was propagated through the atmosphere reaching the ground and reflects back to space while the other solar beam was reflected off a cloud layer that is 4 km above the sea level. The solar radiances in both scenarios were calculated and presented near the $CO_2$ absorption band 1575 nm (6310-6380 $cm^{-1}$) in Figure 6. Left panel in Figure 6 represents the radiance reflected off ground (albedo=0.29, short grass) while the right panel represents the radiance reflected of cloud (albedo= thin cloud). Reflectivity values were selected based on the albedo values of various surface introduced by [Oke, 1992] and [Ahrens, 2006].

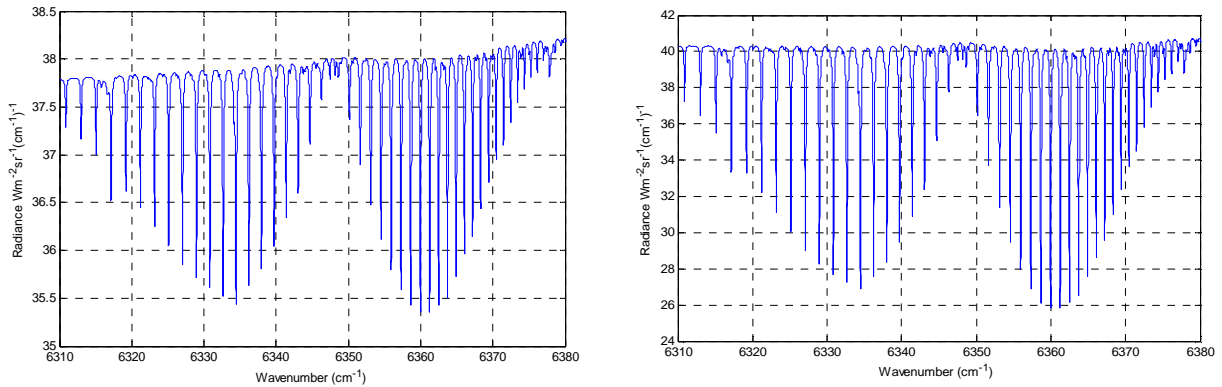

**Figure 6.** Synthetic solar beams reflected off ground (left panel) and off cloud (right panel).

The solar beam that reached the ground and reflected back to space (left panel) yielded approximately 37.78 W m$^{-2}$ sr$^{-1}$ (cm$^{-1}$)$^{-1}$ while the solar beam that reflected off the cloud layer yielded about 40.27 W m$^{-2}$ sr$^{-1}$ (cm$^{-1}$)$^{-1}$. The ratio between the beams reflected off ground and cloud is approximately 1.065 (6.6%). Solar radiance that reaches the ground travels through long atmospheric path and experience more energy loss (attenuation) while the solar beam reflected off cloud travels through short atmospheric path and experience less energy reduction provided that there are no multiple paths above the cloud layer.

4.2 Actual Effect of Clouds

Ratio between cloudy and cloud-free sky conditions has some implications in terms of the $CO_2$ absorption amount. Figure 7 shows the three different cases from three different datasets where Argus 1000 had the chance to collected data over both cloudy and cloud-free skies. The cloudy and cloud-free radiance spectra are shown in the first row of Figure 7. The red spectra were collected over clouds and therefore display high solar radiation while the blue spectra were collected over clear-sky conditions and show less radiation as the photons had the chance to go through the full atmospheric path. The second row in the same figure shows the ratio between the normalized cloudy and cloud-free spectra. The ratio plots show that the difference between cloudy and cloud-free spectra near the $CO_2$ absorption band 1575 nm is approximately 5% implying that the $CO_2$ absorption in the absence of clouds is approximately 5% higher than when clouds are present. The 5% is equal to about 20 parts per million (ppm) of $CO_2$ mixing ratio and this pose a real change in the retrieval process hence the importance of excluding cloudy data in the retrieval process. The third row images show the actual cloud cover for the scene that Argus 1000 collected the analyzed data over. It shows that the cloudy observation numbers (red spectra) were reflected off clouds while the cloud-free observation numbers were reflected of ground.

Solar radiation beam can perform different types of path in the Earth's atmosphere leading to either overestimation or underestimation of $CO_2$ absorption. If, for example, the solar beam undergoes multiple reflections beneath the cloud layer, it will lead to an overestimation in the $CO_2$ absorption. Moreover; if the solar radiation reflects off a cloud layer, it will lead to an underestimation of $CO_2$ absorption therefore introducing error in the quantification of the $CO_2$ mixing ratio in the atmosphere [Alsalem, 2016].

Table 4 shows the synthetic and actual radiance values in the presence and absence of clouds. Though it is extremely difficult to exactly model what is happening in real atmosphere, model and actual findings are very close.

**Table 4**. Model and actual data of cloudy and cloud-free radiances.

|  | Cloudy | Cloud-free | Ratio | Ratio % |
|---|---|---|---|---|
| **Theoretical** | 40.27 | 37.78 | 1.065 | ~ 6.6 |
| **Observation** | 0.279 | 0.267 | 1.044 | ~ 4.5 |

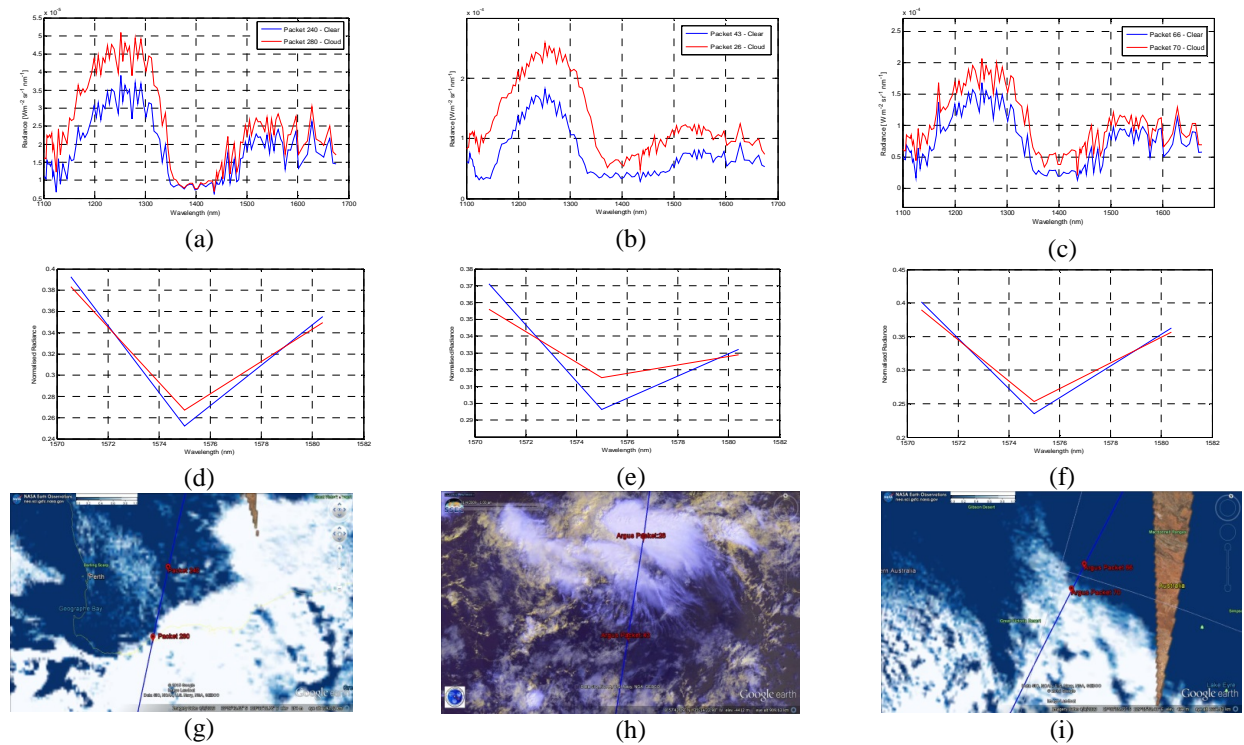

**Figure 7.** The full spectrum of cloudy and clear scenes (first row) for week 41 pass 34, week 9 pass 36 and week 77 pass 28, respectively. Second row shows normalized cloudy and clear spectra at the $CO_2$ band 1575 nm. The third row shows the actual cloud cover. [Cloud imagery by Reto Stockli, NASA's Earth Observatory http://neo.sci.gsfc.nasa.gov/ ]

One of the most important parameters in this analysis is the surface reflectivity (albedo) as it plays a key role in the solar radiation calculations. Fresh snow, for example, has an albedo close to the clouds albedo and would introduce an uncertainty in the retrieval process when data is collected over snow-covered areas in a cloudy day. Thus, data contaminated by clouds must be classified and excluded to accurately infer the greenhouse gases' concentration in the atmosphere. Rehan *et al*. [2015] developed an algorithm capable of detecting and classifying cloud scenes observed by Argus 1000 spectrometer. They also showed that the Argus 1000 is capable of detecting cloud scenes from space [Siddiqui *et al*, 2016].

## 5 Conclusions

We showed that clouds pose a large uncertainty on the space-based retrieval process of $CO_2$. Model finding shows that the theoretical ratio between cloudy and cloud-free radiances is approximately 6.6% while the actual ratio from Argus 1000 spectrometer data is approximately 4.5%. The actual ratio between cloudy and cloud-free sky conditions implies that the $CO_2$ absorption in the absence of clouds is approximately 4.5% higher than when clouds are present. The 4.5% ratio is equal to approximately 20 ppm of $CO_2$ concentration in the Earth's atmosphere and considered to be a large uncertainty in the retrieval process hence the significant need of excluding data contaminated by clouds to accurately infer the $CO_2$ amount in the atmosphere. There is, in general, less atmospheric column path for photons in the presence of clouds but, perhaps, more multi-path so these effects compete.

**Acknowledgments and Data**

The authors would like to acknowledge the Department of Physics and Astronomy and the Department of Earth and Space Science and Engineering at York University for providing supportive learning environments. Dr. Brendan Quine's research funds were used in order to perform this analysis. The authors would also like to acknowledge Thoth Technology Inc. for providing the GENSPECT algorithm and Argus 1000 spectrometer data.

H2O, CO2 and CH4 by Argus 1000 along with GENSPECT line by line radiative transfer model. arXiv preprint arXiv:1608.05386.